\documentclass[]{article}
\usepackage{emulateapj}
\usepackage{graphicx}

\advance \voffset by -0.5cm\relax

\def\msun{{\rm ~M}_{\odot}}

\def\mpy{{\rm ~M}_{\odot} {\rm ~yr}^{-1}}

\begin{document}

\title{Double Compact Objects as Low-frequency Gravitational Wave Sources}

 \author{Krzysztof Belczynski\altaffilmark{1,2}, 
         Matthew Benacquista\altaffilmark{2}, 
         and Tomasz Bulik\altaffilmark{1}}

 \affil{
     $^{1}$ Astronomical Observatory, University of Warsaw, Al.
            Ujazdowskie 4, 00-478 Warsaw, Poland\\
     $^{2}$ Center for Gravitational Wave Astronomy, University of Texas at 
            Brownsville, Brownsville, TX 78520, USA\\
            }

 \begin{abstract} 
We study the Galactic field population of double compact objects (NS-NS, BH-NS, BH-BH 
binaries) to investigate the number (if any) of these systems that can potentially be 
detected with LISA at low gravitational-wave frequencies. We calculate the Galactic 
numbers and physical properties of these binaries and show their relative
contribution from the disk, bulge and halo. 
Although the Galaxy hosts $\sim 10^5$ double compact object binaries emitting
low-frequency gravitational waves, only a handful of these objects in the 
disk will be detectable with LISA, but none from the halo or bulge. This is because 
the bulk of these binaries are NS-NS systems with high eccentricities and
long orbital periods (weeks/months) causing inefficient signal accumulation
(small number of signal bursts at periastron passage in $1$ yr of LISA
observations) rendering them  undetectable in the majority of these cases.
We adopt two evolutionary models that differ in their treatment of the common 
envelope phase that is a major (and still mostly unknown) process in the formation 
of close double compact objects. 
Depending on the adopted evolutionary model, our calculations indicate the likely
detection of about 4 NS-NS binaries and 2 BH-BH systems (model A;  likely survival of 
progenitors through CE) or only a couple of NS-NS binaries 
(model B;  suppression of the double compact object formation due to CE mergers).
\end{abstract}


\section{Introduction}

The {\em Laser Interferometer Space Antenna} (LISA) is a space-based instrument 
to search for and observe gravitational radiation (e.g., Hughes 2006). 
A system of three satellites (5 million kilometers apart), orbiting the Sun, will form 
an interferometer sensitive to low frequency gravitational radiation (GR) 
($\sim 5 \times 10^{-5}-1$ Hz). The main sources of GR at these low frequencies 
are inspirals of super-massive black holes (BHs) in the center of merging galaxies, 
extreme mass ratio inspirals of stellar mass objects into the 
super-massive BHs, and nearby (mostly Galactic) compact binaries.  
Double white dwarf binaries (WD-WD) are the largest population of Galactic systems 
that are expected to produce a confusion-limited noise in the detector, with several 
thousand of the louder systems being potentially resolved. A number of studies have 
concentrated on studies of double white dwarfs in the context of low-frequency 
LISA observations (e.g., Hils, Bender  \& Webbink 1990; Farmer \& Phinney 2003;
Nelemans, Yungelson \& Portegies Zwart 2004; Ruiter et al. 2009, 2010). Here, we 
focus on the other, much less studied group of compact systems; double 
neutron star binaries (NS-NS), black hole-neutron stars (BH-NS), and double black 
holes (BH-BH). Out of all such double compact objects (DCOs) only a handful of NS-NS 
have been discovered in radio surveys (e.g., Lorimer 2005). Although much less common 
than double white dwarfs, DCOs produce much stronger GR 
signals, since they are much denser and more massive than white dwarfs. 

Mergers of all types of DCOs are expected to be the prime 
candidates for high-frequency ground based gravitational-wave interferometers 
like LIGO or VIRGO, while mergers of NS-NS and BH-NS are proposed 
as potential progenitors of short-hard Gamma-ray bursts (Paczynski 1986).
DCOs have been studied extensively over the past 2 decades,
bringing new understanding of their formation, particularly
in the context of population synthesis studies (e.g., Tutukov \& Yungelson 1994; 
Lipunov, Postnov \& Prokhorov 1997; Portegies Zwart \& Yungelson 1998; 
Belczynski \& Bulik 1999; Fryer, Woosley \& Hartmann
1999; Bloom, Sigurdsson \& Pols 1999; Nelemans, Yungelson \& Portegies Zwart 
2001; Hurley, Tout \& Pols 2002; Voss \& Tauris 2003; Dewi \& Pols 2003;
Pfahl, Podsiadlowski \& Rappaport 2005).
Merger rates were recently presented and discussed by Kalogera et al.
(2004; empirical estimates) and Belczynski et al. (2010; population
synthesis).

In this study, we analyze the field Galactic population (disk, bulge and halo)
of DCOs. We do not consider any dynamical
interactions between stars; i.e., evolution of stars in globular clusters is
not accounted for and we evolve only field populations. However, we note
that despite their relative small stellar mass content globular clusters 
may contribute significantly to formation of BH-BH binaries (Kulkarni, Hut \& 
McMillan 1993; Gultekin, Miller 
\& Hamilton 2004;  O'Leary et al. 2006; Sadowski et al. 2008; Downing et al. 2010).  On the other
hand, the formation of DCOs with NSs was found to 
be inefficient in globular cluster environments (e.g., Phinney 1991;
Grindlay, Portegies Zwart, \& McMillan 2006; Ivanova et al. 2008).
Using population synthesis methods we predict numbers 
and physical properties of DCOs, we then calculate their
spatial distribution and estimate the low frequency gravitational
radiation signal that may arise from these binaries.

\section{Modeling}

\subsection{Population Synthesis}
\label{popsynth}

We have used the population synthesis code {\tt StarTrack} to calculate the
numbers and properties of DCOs. The full description of
the code can be found in Belczynski, Kalogera \& Bulik (2002) and Belczynski 
et al. (2008). The code utilizes a set of stellar models (Hurley et al. 2000) 
that allow for evolution of stars at different metallicities. The compact object 
formation follows self-consistently from the stellar models, extended to the 
formation  of the FeNi core (Timmes, Woosley \& Weaver 1996). During the core 
collapse the fall back and direct BH formation is accounted for (Fryer 
\& Kalogera 2001) and the newly born compact objects receive natal kicks (Hobbs 
et al. 2005). Formation of low mass NSs through electron capture 
supernovae is also accounted for (e.g., Podsiadlowski et al. 2004). Binary
interactions are treated in detail and the various processes were calibrated
using either results of detailed evolutionary calculations (e.g., Wellstein,
Langer \& Braun  2001 for mass transfer sequences) or specific sets of 
observations (e.g., Levine, Rappaport \& Zojcheski 2000 for tidal interactions). 

All of our population synthesis calculations implement the standard evolutionary 
model presented in Belczynski et al. (2008). We employ our standard model to
evolve three different  Galactic populations: disk, bulge and halo. 
The disk is assumed to have a stellar mass of $4.0 \times 10^{10} \msun$, stars
have solar metallicity $Z=0.02$ and we assume the age of the population to be 10
Gyr with constant star formation throughout the disk lifetime (i.e., star
formation rate is $4 \mpy$). 
The bulge is assumed to have a stellar mass of $1.1 \times 10^{10} \msun$, stars
also have solar metallicity $Z=0.02$ and we assume the age of the population to be 
10 Gyr with a burst of star formation lasting through the first Gyr of
bulge evolution. 
The halo is assumed to have a stellar mass of $0.1 \times 10^{10} \msun$,  the 
stars have sub-solar metallicity $Z=0.001$ and we assume the age of the population 
to be 13 Gyr with a burst star formation at the very beginning of halo evolution.

Belczynski et al. (2007) noted that many progenitors of DCOs
evolve through a common envelope phase while a donor
star is crossing Hertzsprung gap. Such a star does not have a well developed
core--envelope structure (e.g., Taam \& Sandquist 2000) and once inspiral in a 
common envelope is started 
it may very likely lead to a merger whether there is enough orbital energy
to expel the envelope or not. Taking into account these mergers
leads to a significant decrease in the formation of close DCOs. 
To estimate the impact of this uncertainty on the LISA signal we perform two 
calculations. In one (Model A) we allow for survival in the case of a common 
envelope with a Hertzsprung gap donor (i.e., standard energy balance is
tested to check for system survival; Webbink 1984), while in another (model
B) we assume that all such common envelope events lead to a merger aborting
further binary evolution and thus depleting the population of DCOs.

\subsection{Galactic Model}\label{galmodel}

The binaries are distributed at birth throughout the Galaxy according to stellar 
density models. We assume that the densities are independent of time. For the bulge,
we choose a spherical density with a normal distribution in the radial coordinate and
a cut-off radius of $3.5~{\rm kpc}$, so $\rho_b \propto e^{-(r/r_{0\rm b})^2}$, 
where $r_{0\rm b} = 500~{\rm pc}$ (Nelemans et al. 2004). The disk population is 
assumed to be axially symmetric with cylindrical radius and vertical height 
distribution given by a double exponential 
$\rho_{\rm d} \propto e^{-R/R_{0\rm d}}e^{-|z|/z_{0\rm d}}$,
with $R_{0\rm d} = 2.5~{\rm kpc}$ and $z_{0\rm d} = 200~{\rm pc}$. Finally, we 
distribute the halo binaries according to a simplistic spherical model with 
$\rho_{\rm h} \propto \left(1 + \frac{r}{a_0}\right)^{-3.5}$, with 
$a_0 = 3.5~{\rm kpc}$. The disk and bulge systems are given initial rotational 
velocities in the plane of the disk, and the halo systems are assumed to have 
circular orbits corresponding to their initial positions.
All systems propagate through a Galactic potential and their trajectories 
change due to the kicks received at the birth of each compact object. 

The Galactic gravitational potential is the sum of the 
bulge, disk and halo potentials. The disk and bulge 
 are described  the Miyamoto and Nagai (1975) potentials:
\begin{equation}
\Phi(R,z) = {G M_i \over
\sqrt{ R^2 +(a_i +\sqrt{z^2+b_i}) }}
\end{equation}
where $R=\sqrt{x^2+y^2}$,  the index  $i$ corresponds to bulge and disk, $a_i$ 
and $b_i$ are the parameters, and $M_i$ is the mass. The halo potential is assumed 
to be 
\begin{equation}
\Phi(r) = - {G M_h\over r_c} \left[  {1\over 2}  \ln \left (1+{r^2 \over r_2^2} \right)
+ {r_c\over r}{\rm atan}\left(   {r\over r_c}  \right)          \right] 
\end{equation}
where where $r_c$ is the core radius, and $M_h$ is the parameter describing the 
mass of the halo (Paczynski 1990). The mass of such halo is divergent so we
introduce a cutoff radius $r_{cut}$ beyond which the halo density falls to zero, 
and the potential is $\Phi(r) \propto r^{-1}$.
The bulge is described by  $M_1=1.12\times 10^{10}\msun$,  $a_1 = 0$\,kpc, and 
$b_1=0.277$\,kpc, and the disk by $M_2=8.78\times 10^{10}\msun$, $a_2=4.2$\,kpc, 
and $b_2=0.198$\,kpc.
For the halo potential we use $r_c=6.0$\,kpc, $r_{cut}=100$\,kpc, and $M_h=5.0\times10^{10}\msun$. 
We note that the masses of the potentials differ from the masses of the stellar 
components in Section~\ref{popsynth} because we include non-stellar matter and 
dark matter in the potentials.

\subsection{LISA Signal Simulator}\label{LISASignal}

We calculate the expected signal in LISA as the Michelson signal from an equal-arm 
interferometer using the long wavelength approximation for frequencies below 3 mHz
(Cutler 1998; Benacquista, DeGoes \& Lunder 2004) and the rigid adiabatic approximation 
(Rubbo, Cornish \& Poujade 2004) above this frequency. The gravitational waveform for 
an eccentric binary with angular frequency $\omega=2\pi/P_{\rm orb}$ is calculated in 
the quadrupole approximation (Peters \& Mathews 1963; Peters 1964). We use the specific form 
given in Pierro et al. (2001). The overall amplitude of the gravitational wave is 
proportional to:
\begin{equation}
h_0=\frac{2G^{5/3}\omega^{2/3}M_{\rm chirp}^{5/3}}{rc^4},
\label{gwamp}
\end{equation}
where $r$ is the luminosity distance from the earth to the binary. We assign each binary 
an arbitrary orientation, and so the contribution of each binary to the LISA signal is 
described by 11 parameters. The masses ($M_1$ and $M_2$), luminosity distance ($r$), and 
orbital frequency ($f_{\rm orb}$) are used to construct the overall amplitude and the 
relative amplitudes of the harmonics are obtained from the eccentricity ($e$). The  
sky location is obtained from the final positions ($\theta_s$ and $\phi_s$) of the 
binaries after propagation through the Galactic potential. The arbitrary orientation 
includes the initial phase ($\phi_0$), the argument of the periastron ($\gamma$), and the 
direction angles of the angular momentum vector ($\theta_l$ and $\phi_l$).  We compute 
the one-year time domain signal for each harmonic in the waveform up to a frequency of 30 
mHz. We have not included periastron precession, although this may prove to be important 
for longer observation times.

For $e=0$, all the power in the waveform is concentrated at the $n=2$ harmonic. At 
non-zero eccentricity, the power in this waveform is spread out over several harmonics of 
the orbital frequency. If we define $n_{\rm max}(e)$ to be the harmonic with maximum power 
for eccentricity $e$, then for $e \gtrsim 0.8$, $n_{\rm max}(e)$ can be approximated by
\begin{equation}
n_{\rm max} \simeq \frac{1.6}{(1-e)^{1.5}}.
\end{equation}
The factor of 1.6 arises from maximizing $g(n,e)$ from Peters \& Mathews (1963). Binaries with 
$f_{\rm orb}$ below the low-frequency limit ($f_{\rm crit}$) for 
LISA sensitivity may still be observable by LISA if $n_{\rm max} f_{\rm orb} > f_{\rm crit}$. 
We use the following approximate criterion to separate potential LISA sources from the 
entire DCO population
\begin{equation}
P_{\rm orb} < 1.6 (1-e)^{-1.5} f_{\rm crit}^{-1}
\label{crit1}
\end{equation}
where $f_{\rm crit}=5 \times 10^{-5}~{\rm Hz}$.

LISA instrument noise is simulated by assuming the power spectral density of the noise is 
made up of position (or shot) noise ($S_{\rm np}$) and an acceleration noise ($S_{\rm na}$) 
(converted to strain) given by Cornish (2001). These separate components are combined 
according to:
\begin{equation}
S_{\rm n} = 4 S_{\rm np} + 8S_{\rm na}\left(1+\cos^2{\left(f/f_{*}\right)}\right),
\label{lisanoise}
\end{equation}
where $f_{*} = c/2\pi L$ with the armlength of LISA taken to be $L = 5\times 10^9~{\rm m}$. 
We roll off the acceleration below $f_{\rm crit}$, so that $S_{\rm na}(f \le f_{\rm crit}) 
= S_{\rm na}(f_{\rm crit})$. In reality, the LISA noise will probably not follow this 
simple power law all the way down to our choice of $f_{\rm crit}$, but will begin to rise 
at a higher frequency.

\section{Results}

\subsection{Physical Properties \& Numbers: Model A}

In Table~1 we present the number of DCOs predicted for the
present time in our Galaxy. Additionally, we list the number of potential
LISA systems that satisfy eq.~\ref{crit1} and may produce a GR signal in the frequency
band of LISA. Note that the Galactic population of DCOs is dominated 
by disk ($78\%$) and bulge 
($18\%$) systems with a small contribution of halo binaries ($4\%$). This reflects 
the fact the the number of DCOs is proportional (for a given 
evolutionary model) to the stellar mass. 
Within the entire DCO population, BH-BH systems are dominant 
($67\%$), with a significant contribution of NS-NS binaries ($28\%$) and a
very small fraction of BH-NS systems ($5\%$). The dominance of BH-BH
systems is a consequence of the natal kick model employed, i.e., NSs
are given full kicks, while BHs (due to formation through either partial 
fall back or direct stellar collapse) receive smaller or no natal kicks at all. 

The population of potential LISA sources is quite different than the entire 
DCO population. These are the binaries that were formed at 
close orbits ($P_{\rm orb} \lesssim 150$ day), and constitute only a small 
fraction of the entire population. The NS-NS LISA binaries are $31\%$ of entire
Galactic NS-NS population, while it is $8\%$ for BH-NS and only $1\%$ for BH-BH 
LISA binaries. The decreasing contribution to the LISA population is due to the
fact that many more BH-BH systems are allowed to form on rather wide orbits 
while wide NS-NS binaries are mostly disrupted by natal kicks.
The bias towards disk binaries is even more pronounced for the LISA binaries, in 
that they mostly originate from the disk ($88\%$), with significant contribution 
from the bulge ($11.7\%$) and a negligible number of the systems in halo ($0.3\%$). 
The increased disk contribution is an effect of star formation history. Basically, 
the LISA group consists of close binaries that can merge (and disappear 
from the population) in less than the age of the Galaxy. Since the disk is forming 
stars in an approximately constant manner there is a constant supply of close 
binaries, while for bulge and halo that have formed all the stars $\sim 10$ Gyr 
ago, many close binaries have already merged. 
The most striking feature of LISA population is that it mostly consists of 
NS-NS binaries ($90\%$), with only minor contributions from BH-NS ($4\%$) and
BH-BH ($6\%$) systems. For single star evolution with the adopted IMF it is 
predicted that NSs outnumber BHs by about factor of $5$. Binary evolution further 
affects the numbers. NS-NS progenitors are more likely to avoid 
mergers in RLOF episodes as the two stars in the binary are close in mass, while 
for BH-BH progenitors it is likely that dynamical instability develops and a 
progenitor system enters a common envelope that may lead to a merger. 

\begin{figure}
\includegraphics[width=1.2\columnwidth]{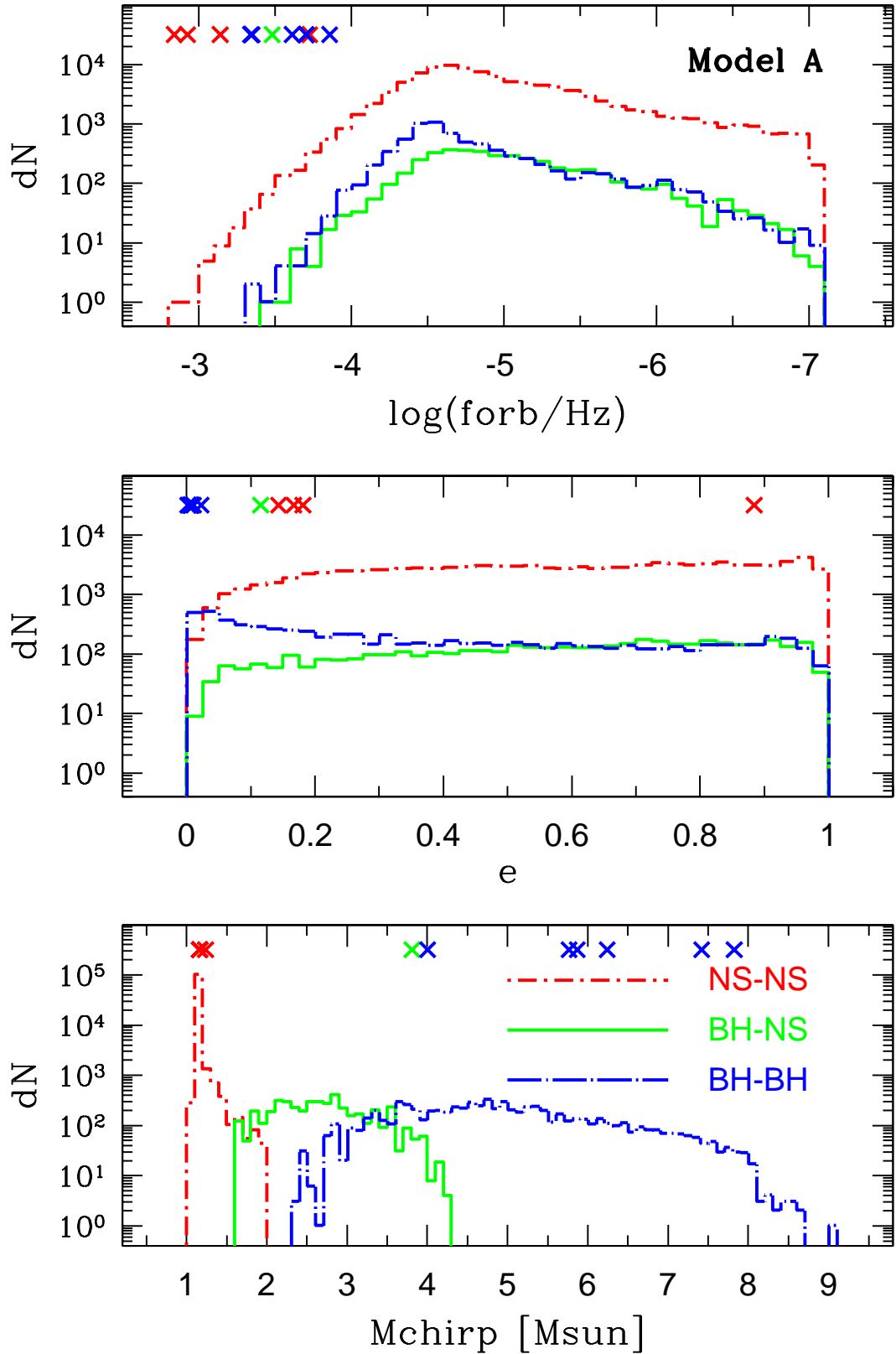}
\caption{Potential LISA Galactic population of double compact objects (Model A). 
The top panel shows the distribution of orbital frequency, middle panel the distribution 
of eccentricities, while the bottom panel has the distribution of chirp masses. Note the  
dominance of NS-NS binaries. 
At the top of each panel we show Galactic resolved double compact objects
for the most optimistic realization in our calculations 
(realization $2$: Table~\ref{resolvedbin}). 
Since this population is very small (4 NS-NS, 1 BH-NS and 6 BH-BH binaries) 
we show all individual resolved sources.
}
\label{pemA}
\end{figure}

In Figure~\ref{pemA} we show the characteristic properties of potential LISA DCOs.  
The orbital frequency ($f_{\rm orb}=1/P_{\rm orb}$) distributions are similar for 
all types of DCOs and they span a wide range: $f_{\rm orb} \sim 10^{-3}-10^{-7}$ Hz
($P_{\rm orb} \sim 0.02 - 150$ day) and peak at $f_{\rm orb} \sim 5 \times 10^{-5}$ 
Hz ($P_{\rm orb} \sim 0.5$ day). For frequencies lower than $f_{\rm crit} = 5 
\times 10^{-5}$ Hz the number of systems drops since the eccentricity of 
a given system needs to increase with decreasing frequency in order for the system 
to be detectable (see eq.~\ref{crit1}). However, the number of systems with higher 
eccentricities is gradually decreasing; for high frequencies
($f_{\rm orb}>f_{\rm crit}$) the systems are so tight ($P_{\rm orb} \lesssim 0.5$d)  
that the orbital decay is very fast (GR emission) and systems merge causing a 
depletion in number with increasing frequency.
The eccentricity distributions are rather flat for all DCOs.
For NS-NS and BH-NS binaries the distributions are slightly skewed 
toward the high $e$-values, while it is just the opposite for BH-BH systems. 
The eccentricity distribution is the direct result of the second supernova
asymmetry. Since systems with NSs receive, on average, larger second kick the
NS-NS and BH-NS binaries are more eccentric than BH-BH systems. The rather  
high fraction ($\sim 40\%$) of BH-BH systems with small eccentricities
($e<0.2$) is either the result of direct collapse of a star to a BH or 
significant fall back of material during the second supernova explosion. 

Chirp mass distributions are very different for the three subclasses. The NS-NS 
chirp mass distribution peaks at $M_{\rm chirp} \sim 1.2 \msun$ with a tail that 
extends to $M_{\rm chirp} \sim 2 \msun$. Systems with BHs have much flatter 
distributions but spanning a wide range of chirp masses; $M_{\rm chirp} \sim 
1.5-4 \msun$ for BH-NS and $M_{\rm chirp} \sim 2.5-9 \msun$ for BH-BH binaries.    
For BH-NS/BH-BH systems the distributions are mostly shaped by the 
mass distribution of BHs. The lowest mass BHs are found just 
over $2.5 \msun$ (adopted maximum NS mass) and they can be as massive as 
$\sim 15 \msun$ for disk and bulge (high metallicity) while they can reach even 
higher masses $\sim 30 \msun$ for halo population (low metallicity). 
Stellar mass BHs are found in our Galaxy with masses up to 
$\sim 15 \msun$ (e.g., Ziolkowski 2010), while in the galaxy IC-10 with 
low metallicity stars a BH was found with mass $\sim 24-33 \msun$ (e.g., 
Prestwich et al. 2007). 
Note that although we predict such high mass BHs in the halo of our
Galaxy there are very few of them, and they do not have very massive
companions as the highest chirp masses for BH-BH binaries are $M_{\rm chirp} 
< 10 \msun$. The system in IC-10 is predicted to form a BH-BH binary with 
 $M_{\rm chirp} \sim 20 \msun$ and a rather short coalescence 
time $\sim 2-3~{\rm Gyr}$ (Bulik, Belczynski \& Prestwich 2008). Therefore, if such a 
system has formed in the halo of our Galaxy it has most probably merged by now.

\subsection{Physical Properties \& Numbers: Model B}

For model B, the most affected systems are BH-NS systems (reduction by
factor of $\sim 2$), and BH-BH binaries (reduction by $\sim 1.4$) while
NS-NS binaries are affected the least (reduction by $\sim 1.2$). All these
changes are not very large especially in the light of the limited prospects for
detecting the majority of these binaries; since the majority of these systems are too
wide to ever make it to the LISA frequency range or to merge within a Hubble time
to be detected by ground-based detectors like LIGO or VIRGO.

\begin{figure}
\includegraphics[width=1.2\columnwidth]{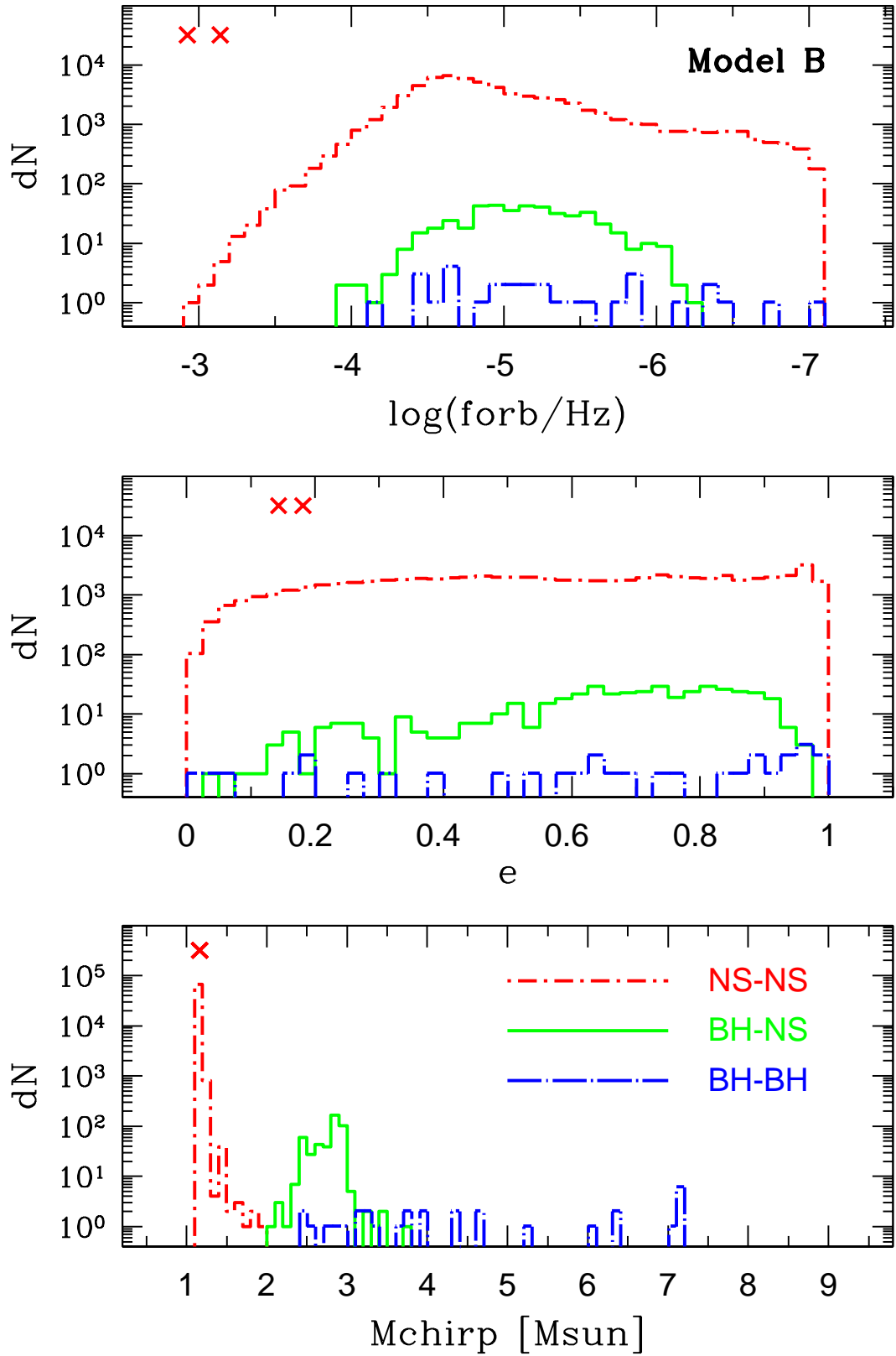}
\caption{Same as Fig.~\ref{pemA} but for model B. The NS-NS dominance is
even more pronounced in this model. Note that there are only 2 resolved sources 
and they are both NS-NS binaries.}
\label{pemB}
\end{figure}

For LISA binaries, the differences are much more pronounced. BH-BH binaries 
are reduced by a factor of $\sim 250$, BH-NS by $\sim 10$, and NS-NS by
 $\sim 1.5$. The progenitors of close LISA binaries are subject to 
one or more common envelope events and thus they are greatly affected in this
model. In particular, progenitors of binaries with BHs that start their
evolution with stars of rather unequal masses are subject to evolving
through the common envelope and merging. Basically, it is predicted that if
there is no survival in a common envelope with Hertzsprung gap donors, then
there are almost no BH-BH and BH-NS binaries in our Galaxy within the LISA
frequency band. In this case, only 
close NS-NS binaries would have a chance to show up in the LISA datastream. 
Although the number of potential LISA NS-NS binaries is reduced, it is still quite
significant ($\sim 7 \times 10^4$). The physical properties of LISA NS-NS 
binaries in Model B are shown in Figure~\ref{pemB} and are not much different 
from Model A.

\subsection{Spatial Distribution}

\begin{figure}
\includegraphics[width=0.9\columnwidth]{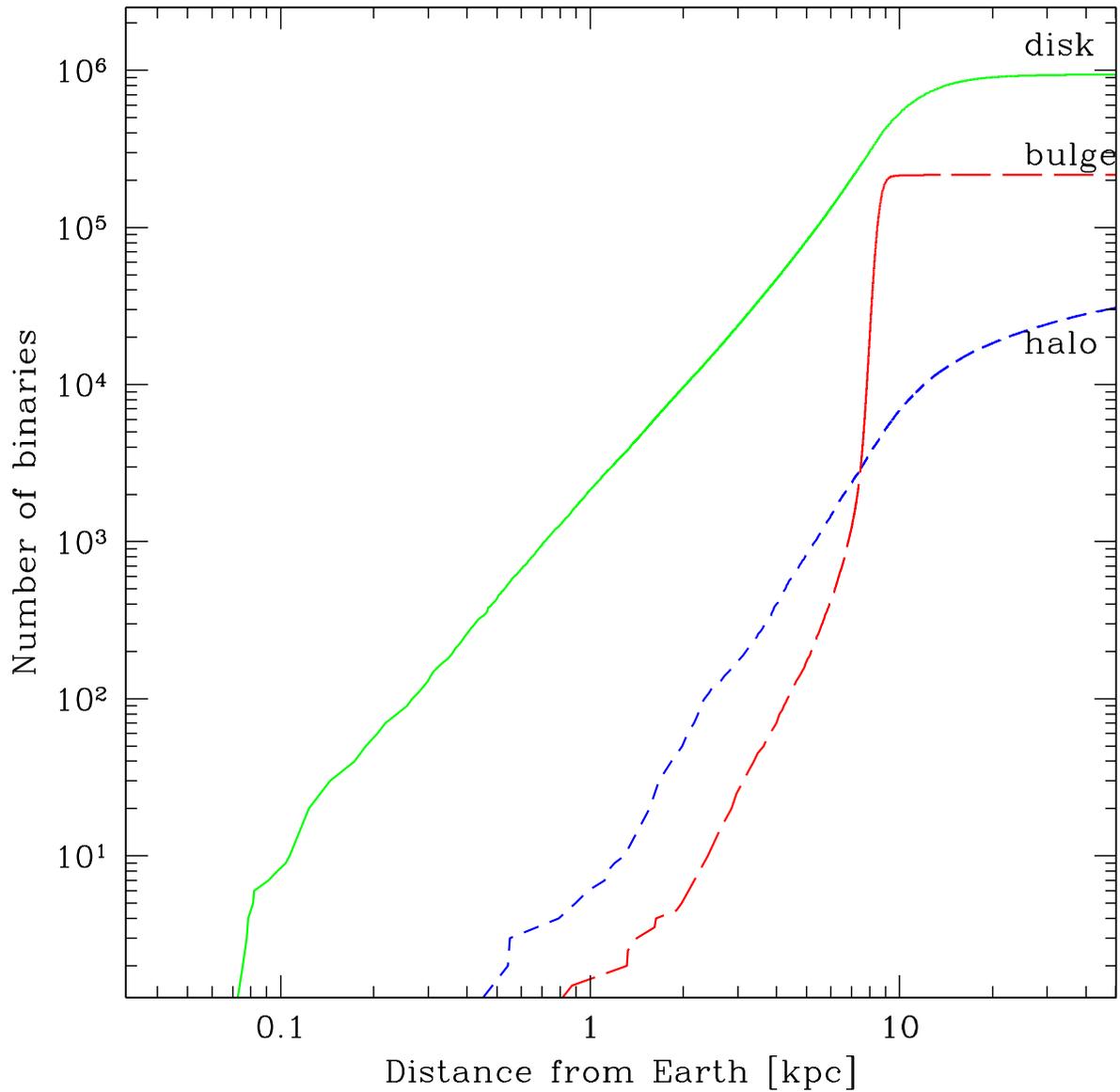}
\caption{The cumulative distribution of the current distances of simulated 
binaries from the Earth (model A). The distances are obtained after propagation 
in the Galactic potential; the binaries have spread out from their birth sites 
due to natal kicks. We show the three Galactic components separately. Note that 
the Galactic center is at a distance of $\sim 8$kpc from Earth.} 
\label{galdyst}
\end{figure}

We present the cumulative distribution (after propagation) of distances of 
DCOs from the Earth in Figure~\ref{galdyst}. 
Halo binaries are born far from the Galactic center and 
their potential energies are large, the kick velocities may increase 
them above zero and unbind some of them. Bulge binaries 
have low potential energies, and adding the kick velocities unbinds only 
a small fraction of them. Their angular momenta are not large since they
are all born close to the Galactic center. Thus they remain concentrated 
in the bulge.
Potential energies of disk binaries have a wide distribution,
depending on the distance from the Galactic center, however, the disk
itself is a potential well. The addition of kick velocities increases
the potential energy. The angular momenta of the disk binaries on their
Galactic orbits are initially aligned perpendicularly to the disk plane, 
yet after the propagation the kick velocities tend to partially isotropize 
them. Thus, they form  a  disk  with scale height of a 
$\approx 0.2$kpc and the radius of  $\approx 15$kpc. The distribution as 
seen from the Earth initially probes this disk for small distances and 
the number of DCOs increases with a cube of distance, yet this
dependence is only seen for the closest few tens of binaries. For
the distances from a few hundred pc to $\approx 10$kpc the number of
DCOs increases roughly as the distance squared because a flat 
distribution of sources is probed. For larger distances ($\gtrsim 15$kpc)  
the dependence flattens out because we reach the end of the disk. The disk
binaries are the most likely potential LISA sources for the entire distance
range in the Galaxy.

\subsection{Gravitational Radiation Signature}

\begin{figure}
\includegraphics[clip=true,width=1\columnwidth]{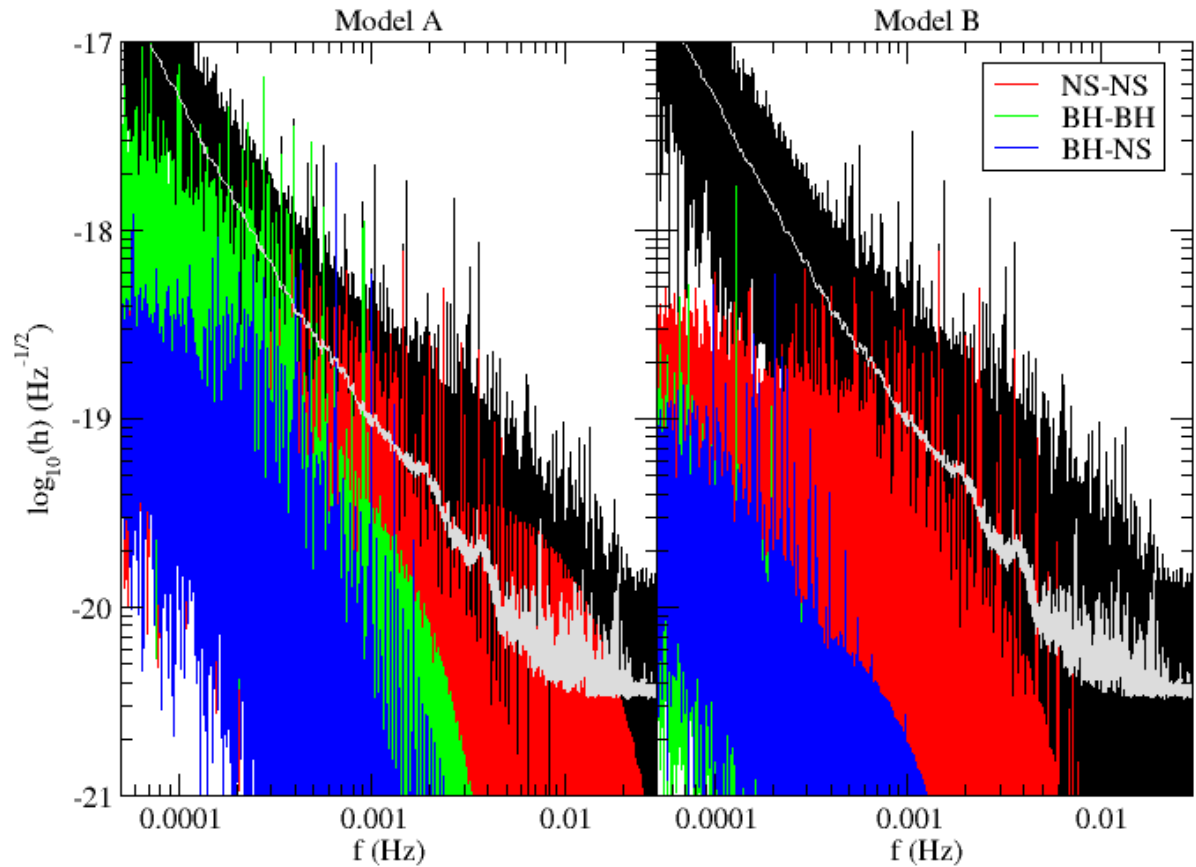}
\caption{Strain spectral densities of the double compact object binaries 
separated by binary type and compared with the full spectrum containing the 
double compact objects, the WD-WD signal and  the LISA noise. The line is a 
running median over the full spectrum. The left panel shows 
model A and the right panel shows model B. Note the strong suppression of the 
BH-BH binaries in model B, and the absence of any detectable signals below 
$\sim 0.4$ mHz.
}
\label{SpecChan}
\end{figure}

\begin{figure}
\includegraphics[width=1\columnwidth]{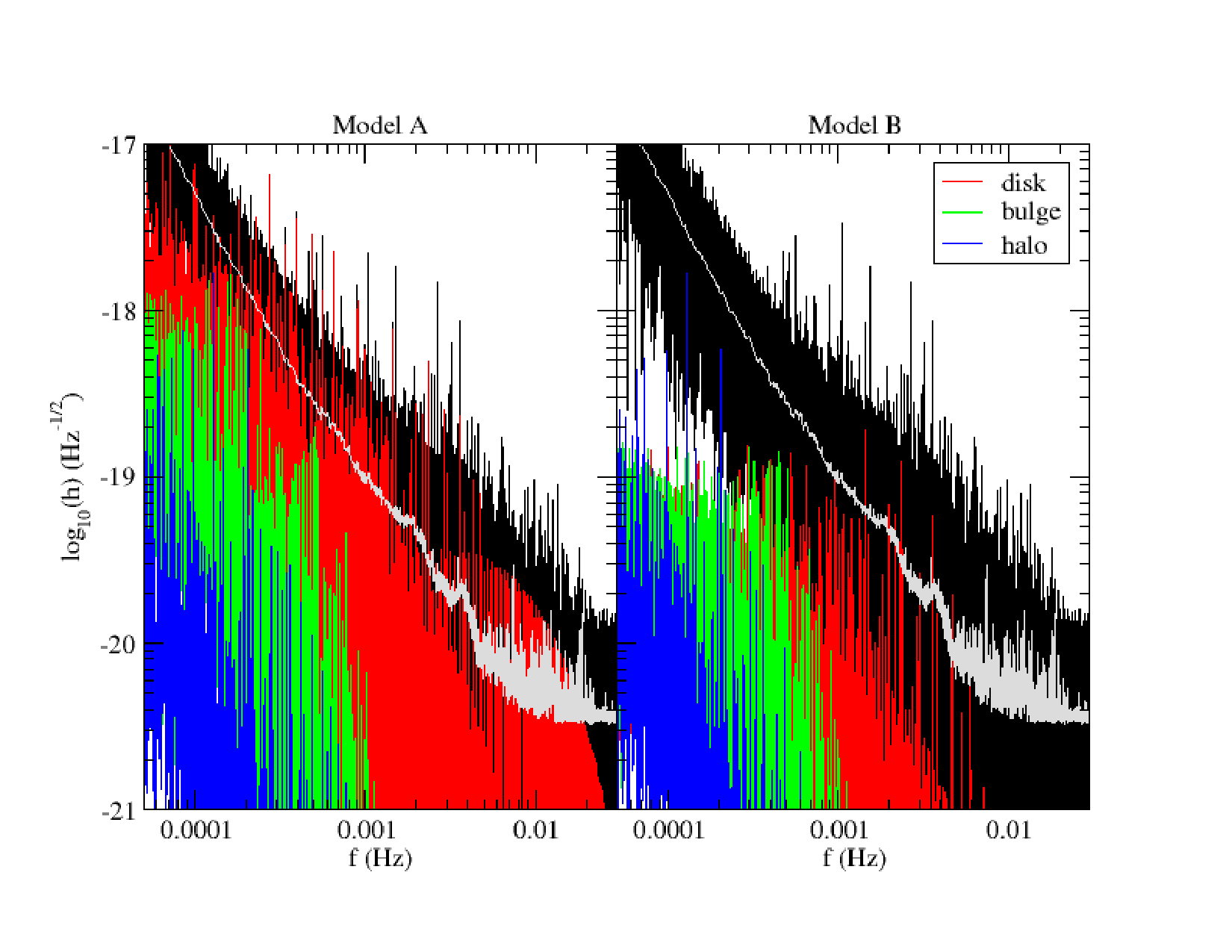}
\caption{Strain spectral densities of the double compact object binaries 
separated by Galaxy component and compared with the full spectrum containing 
the double compact objects, the WD-WD signal and  the LISA noise. The line is 
a running median over the full spectrum. The left panel shows 
model A and the right panel shows model B. Note the dominance of the disk 
binaries in both models, but the significant absence of low frequency binaries 
in model B. 
}
\label{SpecComp}
\end{figure}

The gravitational wave signature for a 1-year observation has been calculated 
for each of 10 realizations of both models. Each realization was obtained by 
redistributing the initial positions of the DCOs according to the stellar 
density distributions described in Section~\ref{galmodel}.
The population has been separated 
into the different binary types in Figure~\ref{SpecChan} and compared with a 
realization of the LISA noise. In Figure~\ref{SpecComp}, we separate out the 
different components of the Galaxy. From this, we see that in both models, the 
disk binaries are the only ones to rise above the LISA noise. 

We are interested in determining if LISA observations can distinguish between 
these two models. Although we do not have an algorithm for identifying eccentric 
stellar mass compact object binaries in the LISA data stream, we can still 
approximate the outcome of such an algorithm. It is possible 
that the coherent signals from the many harmonics of eccentric binaries can be 
summed to boost a signal above the LISA noise even though each individual 
harmonic lies buried in the noise (Benacquista 2001, 2002, Larson \& Hellings 
2009). We assume that some sort of matched-filter will be used, and so we can 
describe the detection statistic as the signal-to-noise ratio ($\rho$) defined 
as (Wainstein \& Zubakov, 1962). For the purposes 
of calculating the detection statistic, we use the discrete analog of 
the signal-to-noise ratio given by Schutz (1997):
\begin{equation}
\label{discretesnr}
\rho^2 = \sum_{k=0}^{N-1}{\frac{\left||\tilde{h}_k\right|^2}{S_k}}
\end{equation}
and take $S_k$ to be the running median over 1000 bins of the power spectrum 
of the LISA noise given by Equation~\ref{lisanoise} combined with a total Galactic 
white dwarf binary population generated using StarTrack (Ruiter et al. 2009,
2010) and the DCO population considered in this work.  We calculate $\tilde{h}_k$ for 
each DCO binary using the LISA signal simulator described in 
Section~\ref{LISASignal}.

The foreground noise due to galactic white dwarf binaries is not Gaussian 
(Timpano et al. 2006). Racine 
\& Cutler (2007) have studied the impact that the non-Gaussianity of the WD-WD  
foreground has on the detection of extreme mass ratio inspirals (EMRIs) and 
supermassive black hole inspirals (SMBHs). Their conclusion is that the threshold 
criterion is essentially unchanged after the loudest white dwarf binary signals 
have been removed from the foreground signals. 

Finally, we note that the imposition of a threshold signal-to-noise ratio 
calculated according to Equation~\ref{discretesnr} can result in detection of binaries 
for which no individual harmonic has a signal-to-noise ratio above 1. Matched 
filtering algorithms have already been demonstrated that can successfully detect 
SMBH signals within mock LISA data (Babak et al. 2008) even though the signal never 
exceeds the instrument and white dwarf binary confusion noise at any frequency (see 
Fig 1 of Arnaud et al. 2007).

We impose a threshold of $\rho \ge 10$ in at least one LISA channel as the criterion for detection. 
We find that the number of resolvable double compact 
object binaries ranges from 3 to 11 in model A and from 0 to 4 in model B. 
The bulk of the potentially detectable very low frequency sources in 
model A are BH-BH binaries that are dramatically suppressed in model B. 
This is borne out by the estimates of resolvable binaries where the number 
of resolvable BH-BH binaries drops to zero in model B.
Thus, detection of high-mass, low-frequency binaries can be used to
distinguish between these two models. In model A there is a relatively
large number of BH-BH binaries at low frequencies:  at $10^{-4}$ Hz there is
1 BH-BH for each 10 NS-NS (Fig.~\ref{pemA}), while there is only 1 BH-BH for 
each 500 NS-NS in model B (Fig.~\ref{pemB}). Since
at these frequencies binaries have relatively large orbital separations,
neutron stars that are of much lower mass than black holes do not generate a
detectable GR signal, while massive black hole systems do. For lower
frequencies ($\lesssim 10^{-5}$ Hz) corresponding to yet larger orbital
separations, even the most massive BH-BH binaries do not generate detectable
signal, although they dominate over other double compact objects. The distribution 
of the potentially resolvable binaries for each realization are presented in 
Table~\ref{resolvedbin}.

\section{Summary}

We have calculated, in a self-consistent way, the gravitational radiation
signal of Galactic DCOs and found that several DCOs can be detected and 
resolved in a 1-year observation with an instrument like LISA. In model A, we 
find comparable numbers of BH-BH and NS-NS binaries, while no BH-BH binaries 
are observed in model B. Such an observation would shed new light
on DCO studies as at present we have only detected 9 NS-NS systems. DCOs with 
a BH are yet to be discovered. LISA could potentially detect several of these 
binaries. 
Starting with NS-NS systems, these potential detections could answer a number
of questions. Do nearly all NS-NS systems host low mass NSs ($\sim
1.35 \msun$) as it seems to appear in a known sample discovered through
radio-pulsar surveys or is it only some observational bias? Galactic merger
rates seem to appear consistent with the notion that NS-NS are  short GRB
progenitors. But would this be confirmed by a sample detected by LISA?  
Detection of any DCOs with BHs would be a feat in and of itself. Although a 
number of these systems are expected from evolutionary calculations, these are 
usually burdened with rather large uncertainties. Detections of BH-BH binaries 
would provide very valuable tests for evolutionary codes that are used in a
number of studies. Knowledge of BH numbers and their basic properties can put 
direct constraints on the evolution of their progenitors---the most massive stars.
In particular, one such example was presented in our study. Depending on a 
common envelope model the number of BH-BH systems that could be detected with 
LISA vanishes. So just by their presence (or lack of thereof) in the 
signal one could test an evolutionary phase that is virtually undetectable
due to its shortness ($\sim 10^3-10^4$ yr). And it is interesting to note,
that although the common envelope was proposed more than 30 years ago
(Paczynski 1976), there is still a surprising lack of understanding of the 
outcome---and the outcome is crucial for formation of Type Ia supernova 
progenitors (close white dwarf binaries) or any type of X-ray binaries that are currently
observed even in external galaxies by {\em Chandra} or {\em XMM}, just to 
name a few examples.   
Although, DCOs will most likely be discovered first in gravitational 
radiation with advanced ground-based interferometers (e.g., LIGO, VIRGO)
they will be detected only in the inspiral and merger phases, preventing a
measurement of their orbital parameters (e.g., separations and eccentricities).
Additionally, the merger rates are too small to expect detections in the
Galaxy, so LISA can provide a small but unique sample of Galactic DCOs. 

The number of detectable compact object binaries that we found in this analysis 
is somewhat smaller than the 42 detectable systems predicted by Nelemans et al. 
(2001). However, this can be reasonably explained by the differences in the 
overall population numbers between the two simulations. The total number of 
compact object binaries predicted in our simulation ranges between  $0.9 \times 
10^6$ for model B and $1.2\times10^6$ for model A, while Nelemans et al. predict 
$4.0 \times 10^6$ such systems. Thus, our prediction of $\sim 11$ systems is 
comparable with Nelemans et al. It should be noted, however, that our estimate 
is conservative as we have assumed no prior removal of any WD-WD signals 
from the calculated value of $S_k$, and that we have assumed an observation time 
of 1 year. In reality, we expect some removal of the loud WD-WD signals and the 
nominal mission lifetime of LISA is 3 years. On the other hand, we have not 
included any penalties incurred by the larger parameter space needed for the 
analysis of eccentric systems.

\section*{Acknowledgments}
We would like to thank Joe Romano, Ashley Ruiter and two referees for a number 
of helpful comments. 
Authors acknowledge the hospitality of the Aspen Center for Physics and 
support from MSHE grants N N203 302835 (TB, KB) and  N N203 404939 (KB) and 
NASA Grant NNX08AB74G (MJB).

\clearpage

\begin{deluxetable}{lccc}
\tablewidth{310pt}
\tablecaption{Number of Double Compact Binaries in the Galaxy\tablenotemark{a}}
\tablehead{         &     NS-NS        &      BH-NS   &   BH-BH       }
\startdata
ALL: disk           & 281091 -- 238245  & 45075 -- 17729 & 621812 -- 424501 \\
ALL: bulge          & 58658 -- 52520    & 10347 -- 4388  & 147260 -- 106170 \\         
ALL: halo           & 6351 -- 6342      & 2845 -- 2776   & 37681 -- 36373   \\
ALL: total          & 346100 -- 297107  & 58267 -- 24893 & 806753 -- 567044 \\
                    &                   &                &                  \\
LISA: disk          & 92253 -- 58633    & 4029 -- 371    & 6903 -- 6        \\
LISA: bulge         & 12853 -- 8700     & 382 -- 48      & 417 -- 0         \\         
LISA: halo          & 298 -- 291        & 50 -- 39       & 36 -- 25         \\
LISA: total         & 105404 -- 67624   & 4461 -- 458    & 7356 -- 31       \\
\enddata
\label{numall}
\tablenotetext{a}{Numbers are given for two models: Model A -- Model B.
ALL -- all binaries in a given category. LISA -- binaries with orbital periods 
and eccentricities that are likely to contribute to GR signal in LISA frequency 
band.} 
\end{deluxetable}

\begin{deluxetable}{ccccccc}
\tablewidth{400pt}
\tablecaption{Distribution of Resolvable Double Compact Binaries in the
Galaxy\tablenotemark{b}}
\tablehead{Realization &  & Model A & & & Model B & \\
& BH-BH & BH-NS & NS-NS & BH-BH & BH-NS & NS-NS  }
\startdata
0 & 3 & 0 & 5 & 0 & 0 & 1 \\
1 & 2 & 0 & 4 & 0 & 0 & 1 \\
2 & 6 & 1 & 4 & 0 & 0 & 2 \\
3 & 2 & 0 & 4 & 0 & 0 & 1 \\
4 & 1 & 1 & 5 & 0 & 0 & 3 \\
5 & 1 & 0 & 2 & 0 & 0 & 0 \\
6 & 1 & 0 & 6 & 0 & 0 & 2 \\
7 & 2 & 0 & 2 & 0 & 0 & 0 \\
8 & 3 & 0 & 3 & 0 & 0 & 2 \\
9 & 2 & 0 & 5 & 0 & 0 & 4 \\
 &  &  &  &  &  &  \\
 & $2.3\pm 1.5$ & $0.2\pm 0.4$ & $4\pm1.3$ &0 & 0 & $1.7\pm1.3$\\
Totals &  & $6.5\pm2.2$ &  &   & $1.7\pm1.3$& 
\enddata
\label{resolvedbin}
\tablenotetext{b}{These numbers represent a conservative estimate as our population 
synthesis model assumes  a stellar disk content of $\sim 4 \times 10^{10}~\msun$, 
while our dynamical calculations use a disk mass content of $\sim 8 \times 10^{10}~\msun$. 
The gas fraction of the disk is around 25\% (Naab \& Ostriker, 2006), leaving the 
remaining $\sim 2 \times 10^{10}~\msun$ as dark matter. If the dark matter content of 
the disk is not this high, then we have underestimated the stellar content of the disk. 
At most, these numbers could be 50\% higher, if there is no dark matter content to the disk.
}
\end{deluxetable}

\clearpage

\clearpage

\clearpage

\clearpage

\begin{references}

\reference{} Arnaud, K., et al.\ 2007, Classical and Quantum Gravity, 24, S551
\reference{} Babak, S., et al.\ 2008, Classical and Quantum Gravity, 25, 114037
\reference{} Benacquista, M., DeGoes, J., \& Lunder, D.\ 2004, CQG, 21, 509
\reference{} Benacquista, M.\ 2002, Classical and Quantum Gravity, 19, 1297
\reference{} Benacquista, M.\ 2001, 20th Texas Symposium, 586, 793
\reference{} Belczynski \& Bulik 1999, \aap, 346, 91
\reference{} Belczynski, K., Kalogera, V., \& Bulik, T.\ 2002a, \apj, 572,
             407
\reference{} Belczynski, K., Taam, R., Kalogera, V., Rasio, F., \& Bulik,
             T.\ 2007, \apj, 662, 504
\reference{} Belczynski, K., et al.\ 2008a, \apjs, 174, 223
\reference{} Belczynski, K., et al.\ 2010, \apj, 715, L138
\reference{} Bulik, T., Belczynski, K., \& Prestwich, A.\ 2008, \apj,
             submitted (arXiv:0803.3516)
\reference{} Bloom, J., Sigurdsson, S., \& Pols, O.\ 1999, MNRAS, 305, 763
\reference{} Cornish, N.J.\ 2001, \prd, 65, 022004
\reference{} Cutler, C.\ 1998, \prd, 57, 7089
\reference{} Dewi, J.D.M., \& Pols, O.R.\ 2003, \mnras, 344, 629
\reference{} Downing, J.M.B., Benacquista, M.J, Giersz, M., \& Spurzem, R., 2010, MNRAS, 407, 1946
\reference{} Farmer, A.J., \& Phinney, E.S.\ 2003, \mnras, 346, 1197
\reference{} Fryer, C., Woosley, S., \& Hartmann, D.\ 1999, ApJ, 526, 152
\reference{} Fryer, C., \& Kalogera, V.\ 2001, \apj, 554, 548
\reference{} Grindlay, J., Portegies Zwart, S., McMillan, S. 2006, Nature,
             2, 116
\reference{} Gultekin, K., Miller, M., \& Hamilton, D.\ 2004, \apj, 616, 221
\reference{} Hils, D., Bender, P.L., \& Webbink R.F.\ 1990, \apj, 360, 75
\reference{} Hobbs, G., Lorimer, D.R., Lyne, A.G., \& Kramer, M.\ 2005,
             \mnras, 360, 974
\reference{} Hughes, S.\ 2006, 6th International LISA Symposium, 873, 13
\reference{} Hurley, J.~R., Pols, O.~R., \& Tout, C.~A.\ 2000, \mnras, 315, 543
\reference{} Hurley, J.\ R.,  Tout, C.\ A., \& Pols, O.\ R.\ 2002, \mnras,
             329, 897
\reference{} Ivanova, N., Heinke, C., Rasio, F., Belczynski, K., \&  Fregeau,
             J.\ 2008, \mnras, 386, 553 
\reference{} Kalogera, V., et al.\ 2004, \apj, 601, L179
\reference{} Kulkarni, S., Hut, P., \& McMillan, S.\ 1993, Nature, 364, 421
\reference{} Larson, S., \& Hellings, R.\ 2009,\baas, 41, 340.
\reference{} Levine, A., Rappaport, S.A., \& Zojcheski, G.\ 2000, 541, L194
\reference{} Lipunov, V., Postnov, K., \& Prokhorov, M.\ 1997, MNRAS, 288, 245
\reference{} Lorimer, D.\ 2005, Living Rev. Relativity, 8, 7
\reference{} Naab, T.\ \& Ostriker, J.P., 2006, MNRAS, 366, 899
\reference{} Nelemans, G., Yungelson, L., \& Portegies Zwart, S.\ 2001,
             \aap, 375, 890
\reference{} Nelemans, G., Yungelson, L., \& Portegies Zwart S.\ 2004,
             \mnras, 349, 181
\reference{} O'Leary, R. M., Rasio, F. A., Fregeau, J. M., Ivanova, N.,
             O'Shaughnessy, R. 2006, \apj, 637, 937
\reference{} Paczynski, B. 1976, in Structure and Evolution of Close Binary Systems;
             Proceedings of the IAU Symposium No. 73, eds. P. Eggleton et al., 1976., 
             p.75
\reference{} Paczynski, B.\ 1986, \apj, 308, L43
\reference{} Paczynski, B.\ 1990, \apj, 348, 485
\reference{} Peters, P., \& Mathews, J.\ 1963, Phys. Rev., 131, 434
\reference{} Peters, P.\ 1964, Phys. Rev. 136, 1224
\reference{} Pfahl, E., Podsiadlowski, P., \& Rappaport, S.\ 2005, ApJ, 628, 343
\reference{} Phinney, S.\ 1991, \apj, 380, L17
\reference{} Pierro, V., Pinto, I., Spallicci, A., Laserra, E., \& Recano, F.\ 
             2001, \mnras, 325, 358
\reference{} Podsiadlowski, P., et al.\ 2004, \apj, 612, 1044
\reference{} Portegies Zwart, S., \& Yungelson, L.\ 1998, \aap, 332, 173
\reference{} Prestwich, A., et al.\ 2007, \apj, 669, L21
\reference{} Racine, E.\ \& Cutler, C.\ 2007, \prd, 76, 124033
\reference{} Rubbo, L., Cornish, N., \& Poujade, O.\ 2004 \prd, 69, 082003
\reference{} Ruiter, A., Belczynski, K., Benacquista, M., \& Holley-Bockelmann, 
             K.\ 2009, \apj, 693, 383  
\reference{} Ruiter, A., Belczynski, K., Benacquista, M., Larson, S., 
             Williams, G.\ 2010, \apj, 717, 1006
\reference{} Sadowski, A., et al.\ 2008, \apj, 676, 1162
\reference{} Schutz, B.\ 1997, Fundamental Physics in Space, ed. 
             A. Wilson, ESA Publication, p. 265
\reference{} Taam, R. E., \& Sandquist, E. L. 2000, \araa, 38, 113
\reference{} Timmes, F.X., Woosley, S.E., \& Weaver, T.A.\ 1996, \apj, 457, 834
\reference{} Timpano, S.E., Rubbo, L.J., \& Cornish, N.J.\ 2006, \prd, 73, 122001
\reference{} Tutukov, A., \& Yungelson, L.\ 1994, MNRAS, 268, 871
\reference{} Vecchio, A. \& Wickham, E.D.L.,\ 2004, \prd, 70, 082002
\reference{} Voss, R., \& Tauris, T.\ 2003, MNRAS, 342, 1169 
\reference{} Wainstein, L.A.\ \& Zubakov, V.D.\ 1962, ``Extraction of Signals 
             from Noise'', Prentice-Hall
\reference{} Webbink, R.\ 1984, \apj, 277, 355
\reference{} Wellstein, S., Langer, N., \& Braun, H.\ 2001, \aap, 369, 939
\reference{} Ziolkowski, J.\ 2010, MmSAI, 81, 294 


\end{references}
\end{document}